\documentclass{iau307}
\usepackage{graphicx}
\usepackage{natbib}
\usepackage{url}
\usepackage{dtklogos}
\bibpunct{(}{)}{;}{a}{}{,}

\title[Cepheid Period Cepheid \& Stellar Evolution] 
{Pulsation Period Change \& Classical Cepheids: Probing the Details of Stellar Evolution}

\author[Neilson et al.]   
{Hilding R.~Neilson$^1$,
Alexandra~C.~Bisol$^2$,
 Ed Guinan$^2$,
 \and Scott Engle$^2$
 }

\affiliation{$^1$ East Tennessee State University \\ email: {\tt neilsonh@etsu.edu} \\[\affilskip]
$^2$Villanova University}

\pubyear{2014}
\volume{307} 
\pagerange{}
\jname{New windows on massive stars: asteroseismology, interferometry, and spectropolarimetry}
\editors{G. Meynet, C. Georgy, J.H. Groh \& Ph. Stee, eds.}

\begin{document}

\maketitle

\begin{abstract}
Measurements of secular period change probe real-time stellar evolution of classical Cepheids making these measurements powerful constraints for stellar evolution models, especially when coupled with interferometric measurements. In this work, we present stellar evolution models
and measured rates of period change for two Galactic Cepheids: Polaris and \emph{l}~ Carinae, both important Cepheids for anchoring the Cepheid Leavitt law (period-luminosity relation). The combination of previously-measured parallaxes, interferometric angular diameters and rates of period change allows for predictions of Cepheid mass loss and stellar mass. Using the stellar evolution models, We find that \emph{l}~Car has a mass of about 9~$M_\odot$ consistent with stellar pulsation models, but is not undergoing enhanced stellar mass loss. Conversely, the rate of period change for Polaris requires including enhanced mass-loss rates. We discuss what these different results imply for Cepheid evolution and the mass-loss mechanism on the Cepheid instability strip.

\keywords{stars: individual (\emph{l}~Carinae, Polaris), stars: mass loss, Cepheids}
\end{abstract}

\firstsection 
\section{Introduction}
Classical Cepheids are powerful probes of stellar structure thanks to their pulsation periods. Measurements of the change of pulsation period directly constrain stellar evolution models \citep{Eddington1919}. This has consequences for understanding the transition from hot, blue main sequence stars to the end points of asymptotic giant branch evolution and supernovae progenitors.
But, measuring the rate of period change requires decades of time-domain observations, spanning about one centur. For \emph{l}~Carinae, $P = 35.5$~days, and Polaris, $P = 4.97$~days, we measure $\dot{P} = 4.46\pm 1.46$ and $23.7\pm 6.5$~s~yr$^{-1}$, respectively.

\section{Rates of Period Change}
\begin{figure}[t]
\begin{center}
\includegraphics[width=0.4\textwidth]{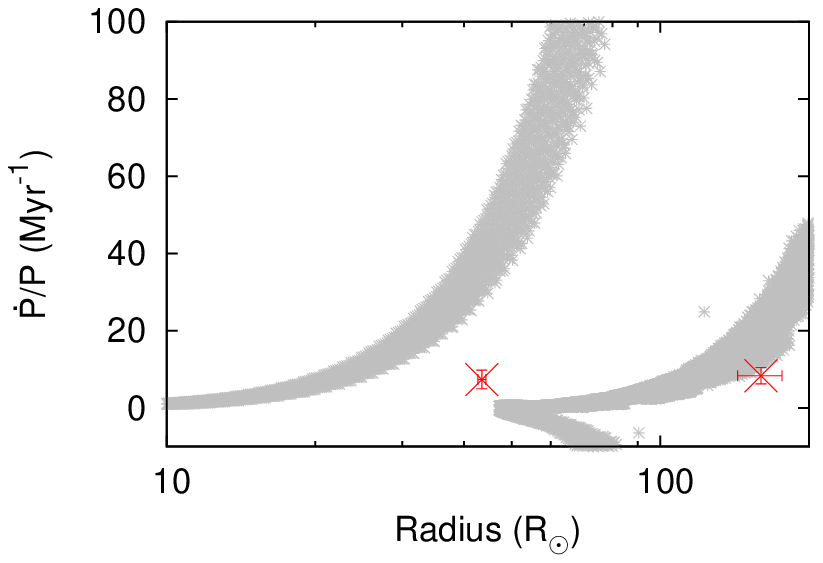} \includegraphics[width=0.4\textwidth]{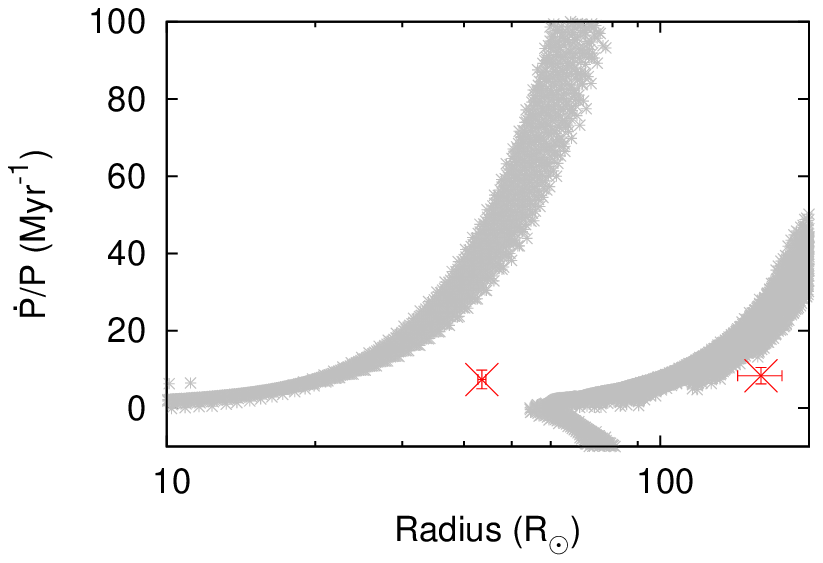} 
\caption{Rates of period change relative to pulsation period from stellar evolution models of Cepheids assuming $10^{-9}$ (left) and $10^{-6}~M_\odot~$yr$^{-1}$ (right)
during the Cepheid phase of evolution. The red points represent measured rates for Polaris and \emph{l}~Car. There are three sequences of period change, large positive correspond to the first crossing of the instability strip while the smaller positive sequence is the third crossing. Negative rates represent the second crossing.}
\label{fig1}
\end{center}
\end{figure}

Period change is measured from time-series observations of the Cepheid light curve and computing an O-C diagram that plots the period measured at some time minus a reference period. A parabolic structure indicates that the period is changing and that change is constant \citep{Percy2007}. However, determining the evolutionary state of Cepheid from period change is not obvious as there are three crossings of the instability strip. Two crossings occur when the star is evolving red ward and period change is positive. The second crossing from the red giant stars to hotter effective temperatures corresponds to negative rates of period change. Hence, we require another constraint, one provided by interferometry. For instance, assuming parallaxes from {\it Hipparcos} and HST \citep{Leeuwen2007, Benedict2007}, the radius of Polaris and \emph{l}~Car are $43.5 \pm 0.8$ and $159.9 \pm16.6~R_\odot$, respectively \citep{Kervella2006, Merand2006}. This permits the comparison of period changes and radii for these stars to constrain evolution and fundamental properties. We plot synthetic rates of period change in Fig.~\ref{fig1} based on stellar evolution models computed using the Bonn code \citep{Neilson2011,Neilson2012a,Neilson2012b} for two cases: the first assuming a Cepheid mass- loss rate of $10^{-9}$ and $10^{-6}~M_\odot~$yr$^{-1}$.

\section{Outlook}

Based on the models, Polaris is poorly represented. We suggest that Polaris is undergoing enhanced mass loss at the rate of order $10^{-7}$ - $10^{-6}~M_\odot~$yr$^{-1}$, but stellar evolution models do not evolve across the instability strip at those rates making it challenging to reproduce the observed properties \citep{Neilson2014}.
The period change for \emph{l}~Car is consistent only with stellar evolution models with smaller mass-loss rates $ < 10^{-7}~M_\odot~$yr$^{-1}$.

Based on that comparison, we determine fundamental parameters such as mass, $M = 8.6\pm 0.5~M_\odot$, effective temperature $T_{\rm{eff}} = 4960\pm 280~$K and  luminosity $\log L/L_\odot = 4.04\pm 0.09$. These results provide valuable insights into the evolution of stars at the mass threshold between core-collapse supernovae and white dwarf formation. Interestingly, the findings hint at a possibility that, counterintuitively, massive Cepheids have weaker winds than lower-mass Cepheids. This conjecture is based on the combination of time-domain astronomy and optical interferometric observations constraining stellar evolution models.

\acknowledgements
We acknowledge funding from NSF grants AST-0807664 and  AST-0507542 and NASA grants GO-12302 and GO-13019.

\bibliographystyle{iau307}
\bibliography{proc}

\end{document}